\documentstyle[12pt]{article}
\begin{document}

\begin{center}

{\bf \LARGE Light-front two-dimensional QED:\\
            Self-field approach\\}

\vspace{2 cm}

{\bf \Large Fuad M. Saradzhev}

\vspace{1 cm}

{\small \it Institute of Physics, Academy of Sciences of Azerbaijan,\\
            Huseyn Javid pr. 33, 370143 Baku, AZERBAIJAN\\}

\vspace{1 cm}

{\bf Abstract}

\end{center}

The self-field approach to quantum electrodynamics  $(QED)$
is used to study the bound state problem in light-front
two-dimensional $QED$ with massive matter fields. A composite
matter field describing bound states is introduced and the
relativistic bound state equation for the composite field
including a self-potential is obtained. The Hamiltonian form
of the bound state equation in terms of the invariant mass
squared operator is given. The eigenvalue problem of this
operator is solved for a fixed value of the self-potential,
the corresponding eigenfunctions and the mass spectrum
are found. In the case of massless matter fields , there are
no self-field terms in the bound state equation, and the
invariant mass spectrum can be evaluated explicitly. Possible
ways of deriving more complete information about the bound
state spectrum are briefly discussed.

\newpage

\begin{center}

\section{INTRODUCTION}

\end{center}

Bound state problems in relativistic field theory, particularly
quantum electrodynamics $(QED)$, are basic ones for particle
physics. It is well known that in perturbative $QED$ the bound
state problems cannot be treated starting from first principles.
Instead one begins from a Shrodinger or Dirac-like equation
obtained from some approximation to the Bethe-Salpeter relations
and then calculates the perturbation diagramms to the bound state
solutions of these equations.

A nonperturbative treatment of two and many body systems in closed
form is possible in the self-field formulation of $QED$
\cite{kraus83,barut90}. Here one starts from two fermion fields
${\psi}_1$ and ${\psi}_2$ coupled by the usual electromagnetic
minimal coupling. One then eliminates the electromagnetic field,
introduces a composite field ${\Phi} = {\psi}_1 \otimes {\psi}_2$
and derives a $2$-body wave equation for this composite field
including the radiative corrections. This relativistic bound state
equation is a genuine and exact one obtained directly from the
action without using any approximation. For two-dimensional $QED$,
the $2$-body wave equation was constructed and analyzed in
\cite{barsa94}.

In the present paper we aim to apply the self-field approach to
light-front two-dimensional $QED$ $(QED_2)$ defined
on the circle. Light-front
quantization \cite{brod91}
has a number of advantages. These include kinematical Lorentz
boosts and a simpler vacuum structure
. By quantizing at equal time on the light-front
a gauge theory can be reduced to an eigenvalue problem for
the
invariant
mass squared operator. The relativistic spectrum emerges as the set of
eigenvalues of this operator.

In the self-field approach the quantum theory is first quantized
. The  electromagnetic field has no separate degrees of freedom,
they are determined by the matter degrees of freedom, but then
one must include nonlinear self-field terms. We consider the
standard $QED_2$ with one matter field $\psi$ and describe bound
states by the composite field ${\Phi}={\psi} \otimes {\psi}$.
We use the advantages of the light-front formulation in order
to get exact expressions for the bound states wavefunctions
and spectrums for both massless and massive fermions.

Our paper is organized as follows. In Sec. 2 we give the light-
front formulation of $QED_2$. In Sec. 3 we introduce the
composite matter field and rewrite the action of the model entirely
in terms of this field. We then vary the action with respect to
the composite field and derive a bound state equation. In Sec. 4
we give the Hamiltonian form of the bound state equation in
terms of the invariant mass squared operator $M^2$. We solve
exactly the eigenvalue problem for $M^2$ for a fixed value of
the self-potential and find explicitly the corresponding
eigenfunctions and the spectrum. We consider also the massless
case when mass  of the matter fields is zero. In Sec. 5 we
conclude with discussion.

\newcommand{\ren}{\renewcommand{\theequation}{2.\arabic{equation}}}
\newcommand{\new}[1]{\renewcommand{\theequation}{2.\arabic{equation}#1}}
\newcommand{\add}{\addtocounter{equation}{-1}}

\begin{center}

\section{LIGHT-FRONT $2$-DIM QED}

\end{center}

The two-dimensional quantum electrodynamics describes matter
fields interacting with an electromagnetic field in two-
dimensional space-time. The Lagrangian density of $QED_2$ is
\ren
\begin{equation}
{\cal L} = i \bar{\psi} {\gamma}^{\mu} {\partial}_{\mu} {\psi}
- m \bar{\psi} {\psi} - \frac{1}{4} F_{\mu \nu} F^{\mu \nu} -
e \bar{\psi} {\gamma}^{\mu} {\psi} A_{\mu} ,
\end{equation}
where $(\mu,\nu) = \overline{0,1}$ , ${\gamma}^{\mu}$ are Dirac
matrices, $F^{\mu \nu} \equiv {\partial}^{\mu} A^{\nu} -
{\partial}^{\nu} A^{\mu}$ is the electromagnetic field strength.
The matter field ${\psi}$ is $2$-component Dirac spinor, and
$\bar{\psi} = {\psi}^{\star} {\gamma}^0$.

We choose the light-front coordinates $x^{\pm} = x^0 \pm x^1$ ,
$x^{+}$ and $x^{-}$ playing the role of time and space
coordinates respectively. The metric tensor $g^{\mu \nu}$ for
the light-front coordinates has the form
\[
g^{++} = g^{--}=0 , \hspace{1 cm}  g^{+-}=g^{-+}=2 .
\]
We must distinguish upper and lower indices; for example, time
and space derivatives are
\[
{\partial}^{-} = 2 {\partial}_{+} \equiv 2 {\partial}/{\partial}x^{+},
\]
\[
{\partial}^{+} = 2 {\partial}_{-} \equiv 2 {\partial}/{\partial}x^{-}.
\]

The algebra of Dirac matrices ${\gamma}^{\pm} = {\gamma}^0 \pm
{\gamma}^1$ is
\[
{\gamma}^{+}{\gamma}^{+} = {\gamma}^{-}{\gamma}^{-} =0,
\]
\[
{\gamma}^{+}{\gamma}^{-} + {\gamma}^{-}{\gamma}^{+} =4.
\]
Using this algebra, we can define projection operators
${\Lambda}^{(\pm)} \equiv \frac{1}{4} {\gamma}^{\mp} {\gamma}^{\pm}$
and projected spinors ${\psi}_{\pm} \equiv {\Lambda}^{(\pm)} {\psi}$.

In terms of the light-front coordinates the Lagrangian density
(2.1) is rewritten as
\[
{\cal L} = ({\psi}_{+}^{\star} i {\partial}^{-} {\psi}_{+} +
{\psi}_{-}^{\star} i {\partial}^{+} {\psi}_{-} ) -
\frac{1}{2} m ({\psi}_{-}^{\star} {\gamma}^{+} {\psi}_{+} +
{\psi}_{+}^{\star} {\gamma}^{-} {\psi}_{-} )
\]
\ren
\begin{equation}
+ \frac{1}{2} ({\partial}_{-} A_{+} - {\partial}_{+} A_{-})^2
- \frac{1}{2} e (j^{+} A_{+} + j^{-} A_{-} ),
\end{equation}
where $A_{\pm} = A_0 \pm A_1$ , while $j^{\pm} \equiv
2{\psi}_{\pm}^{\star} {\psi}_{\pm}$ are light-front matter currents.

 In what follows we work in the light-front gauge $A_{-}=0$.
We    
suppose that light-front space is a circle of length $L$,
$0 \leq x^{-} < L$ , and impose the boundary conditions
\new{a}
\begin{equation}
A_{+}(x^{+},0) =A_{+}(x^{+},L),
\end{equation}
\add
\new{b}
\begin{equation}
{\psi}_{\pm}(x^{+},L) = \exp\{i2{\pi}{\kappa}_{\pm}\}
{\psi}_{\pm}(x^{+},0) ,
\end{equation}
${\kappa}_{\pm}$ being arbitrary numbers.

The action of the light-front $QED_2$ is
\[
W=\int_{-\infty}^{\infty} dx^{+} \int_0^L dx^{-}
{\cal L}(x^{+},x^{-}).
\]

The Lagrange-Euler   equations deduced from it are
\new{a}
\begin{equation}
{\partial}^{+} E = - e j^{+},
\end{equation}
\add
\new{b}
\begin{equation}
{\partial}^{-} E = e j^{-},
\end{equation}
and
\new{a}
\begin{equation}
i{\partial}^{+}{\psi}_{-} =\frac{1}{2} m {\gamma}^{+} {\psi}_{+},
\end{equation}
\add
\new{b}
\begin{equation}
i{\partial}^{-}{\psi}_{+} =\frac{1}{2} m {\gamma}^{-} {\psi}_{-}
+ e A_{+} {\psi}_{+},
\end{equation}
where $E \equiv {\partial}_{-} A_{+}$ is
the electric field strength. From (2.4a-b) we have
\[
{\partial}_{-} j^{-} + {\partial}_{+} j^{+} =0,
\]
i.e. the matter current is conserved.

The equation (2.4a) gives us the Gauss' law and the
boundary conditions for $E(x^{-},x^{+})$ :
\[
E(L,x^{+}) - E(0,x^{+}) = - e\frac{1}{2} \int_0^L dx^{-}
j^{+} \equiv - Q^{+},
\]
i.e., the electric field is not single-valued if the
light-front charge $Q^{+}$ is non-zero (see below).

The equation for ${\psi}_{+}$ involves the light-front
time derivative, so ${\psi}_{+}$ is a dynamical degree
of freedom. On the other hand, the equation for
${\psi}_{-}$ involves only spatial derivative, so
${\psi}_{-}$ is a constrained degree of freedom that
should be eliminated in favour of ${\psi}_{+}$:
\ren
\begin{equation}
{\psi}_{-}(x^{-},x^{+}) = - \frac{i}{8} m {\gamma}^{+}
\int_{0}^{L}  \epsilon(x^{-} - y^{-}) {\psi}_{+}(y^{-},x^{+})
dy^{-} .
\end{equation}
The solution (2.6) fulfils  antiperiodic boundary conditions,
${\psi}_{-}(L,x^{+})=-{\psi}_{-}(0,x^{+})$,
so that ${\kappa}_{-}= \pm \frac{1}{2}$.

The action of the electromagnetic field can be reexpressed by a
partial integration, using (2.4a-b) and the boundary conditions, as
\[
\frac{1}{2} \int_{-\infty}^{\infty} dx^{+} \int_0^L dx^{-}
({\partial}_{-} A_{+} )^2 =
\frac{e}{4} \int_{-\infty}^{\infty} dx^{+} \int_0^L dx^{-}
A_{+} j^{+} - \frac{1}{2} \int_{-\infty}^{\infty} dx^{+}
A_{+}(0,x^{+}) Q^{+}.
\]
With (2.6) , the total action becomes
\[
W[{\psi},A] = \int_{-\infty}^{\infty} dx^{+} \int_0^L dx^{-}
 ({\psi}_{+}^{\star} i {\partial}^{-} {\psi}_{+} 
- \frac{e}{4} j^{+}A_{+} ) - \frac{1}{2} \int_{-\infty}^{\infty}
dx^{+} A_{+}(0,x^{+}) Q^{+}
\]
\ren
\begin{equation}
- \frac{i}{2} m \int_{-\infty}^{\infty} dx^{+} \int_0^L dx^{-}
\int_0^L dy^{-} {\psi}_{+}^{\star}(x^{-},x^{+})
\epsilon(x^{-} - y^{-}) {\psi}_{+}(y^{-},x^{+}).
\end{equation}
If we solve Eqs.(2.4a-b), express $A_{+}$ in terms of $j^{+}$
and insert the expression obtained into Eq.(2.7),
then we get an action written only in terms of the matter fields.

The electromagnetic field equations can be rewritten in the form
\new{a}
\begin{equation}
{\partial}_{-}^2 A_{+} = - \frac{1}{2} e j^{+},
\end{equation}
\add
\new{b}
\begin{equation}
{\partial}_{-}
{\partial}_{+} A_{+} =  \frac{1}{2} e j^{-}.
\end{equation}

Eqs.(2.8a-b) with the periodic boundary conditions 
for the gauge field are 
solved by
\ren
\begin{equation}
A_{+}(x^{-},x^{+}) = - \frac{1}{2} e \int_0^L dy^{-}
D(x^{-},y^{-}|L) j^{+}(x^{+},y^{-}),
\end{equation}
where the Green's function is
\[
D(x^{-},y^{-}|L) \equiv \frac{1}{2} |x^{-} - y^{-}| +
\frac{x^{-}y^{-}}{L} - \frac{1}{2} x^{-} ,
\]
i.e. $A_{+}(x^{-},x^{+})$ is completely determined by
$j^{+}(x^{-},x^{+})$ .

The Green's function $D(x^{-},y^{-}|L)$ is not symmetric
in $x^{-}$ and $y^{-}$. The reason for that is non-zero
$Q^{+}$. We can easily see this if for a moment use
\[
D^{sym}(x^{-},y^{-}|L) \equiv \frac{1}{2} |x^{-} - y^{-}|
+ \frac{x^{-}y^{-}}{L},
\]
\[
D^{sym}(x^{-},y^{-}|L) = D^{sym}(y^{-},x^{-}|L),
\]
instead of $D(x^{-},y^{-}|L)$ in Eq.(2.9).
The boundary conditions for the field $A_{+}$ become
\[
A_{+}(L,x^{+}) = A_{+}(0,x^{+}) + \frac{1}{2} Q^{+},
\]
i.e. $A_{+}$ is periodic only if $Q^{+}=0$. So
$D^{sym}(x^{-},y^{-}|L)$ is a right choice only for the
vanishing charge $Q^{+}$.

However, only the symmetric part of the Green's function
contributes to the action. If we insert (2.9)
in (2.7) , we
obtain
the action in the light-front  gauge as
\[
W[{\psi}] = \int_{-\infty}^{\infty} dx^{+} \int_0^L dx^{-}
{\psi}_{+}^{\star} i {\partial}^{-} {\psi}_{+}
- \frac{i}{2} m \int_{-\infty}^{\infty} dx^{+}
\int_0^L dx^{-} \int_0^L dy^{-} {\psi}_{+}^{\star}(x^{-},x^{+})
\epsilon(x^{-} - y^{-}) {\psi}_{+}(y^{-},x^{+}) 
\]
\begin{equation}
+ \frac{e^2}{8} \int_{-\infty}^{\infty} dx^{+} \int_0^L dx^{-}
\int_0^L dy^{-} j^{+}(x^{-},x^{+}) D^{sym}
(x^{-},y^{-}|L) j^{+}(y^{-},x^{+}),
\end{equation}
the last term representing
the current-
current interaction.

\renewcommand{\ren}{\renewcommand{\theequation}{3.\arabic{equation}}}
\renewcommand{\add}{\addtocounter{equation}{-1}}
\renewcommand{\new}[1]{\renewcommand{\theequation}{3.\arabic{equation}#1}}
\newcommand{\set}{\setcounter{equation}{0}}

\set

\begin{center}

\section{BOUND STATE EQUATION}

\end{center}

Let us define a composite field ${\Phi}$ by
\ren
\begin{equation}
{\Phi}(x_1^{-},x^{+}|x_2^{-},x^{+}) \equiv {\psi}_{+}(x_1^{-},x^{+})
\otimes {\psi}_{+}(x_2^{-}, x^{+}).
\end{equation}
This is a $4$-component spinor field , ${\Phi}^{\alpha \beta} =
{\psi}_{+}^{\alpha} {\psi}_{+}^{\beta}$ , $(\alpha, \beta)=
\overline{1,2}$. However, only one component of the composite
field is nonvanishing:
\[
{\Phi}^{11} ={\psi}_{+}^1 {\psi}_{+}^1 =
{\Phi}^{++} ,
\]
\[
{\Phi}^{12} =
{\Phi}^{21} =
{\Phi}^{22}  =0.
\]
The configuration space $(x_1^{-} , x_2^{-})$ is a square of side $L$
$(0 \leq x_{1}^{-} < L$ , $0 \leq x_2^{-} <L)$ with the opposite sides
being identified, i.e. a torus.

We can rewrite the action (2.10) entirely in terms
of the composite field ${\Phi}$. In order to do this we multiply
the  action with the normalization factor
(which is constant of motion)
\begin{equation}
\int_0^L dx^{-} {\psi}_{+}^{\star}(x^{-},x^{+}) {\psi}_{+}(x^{-}, x^{+})=1.
\end{equation}
The resultant action in terms of the composite field is
\[
W[{\Phi}]= \frac{1}{2} \int_{-\infty}^{\infty} dx^{+} \int_0^L dx_1^{-}
\int_0^L dx_2^{-} {\Phi}^{\star,++}(x_1^{-}, x^{+}|x_2^{-}, x^{+})
 (  {\pi}_{(1)}^{-} +
 {\pi}_{(2)}^{-}
\]
\begin{equation}
- m {\phi}^{self} )
{\Phi}^{++}(x_1^{-}, x^{+}|x_2^{-},x^{+}).
\end{equation}
Here the index $(1)$ refers to the coordinates
of the first field ${\psi}_{+}(x_1^{-}, x^{+})$ in the ansatz
(3.1), the index $(2)$ - to
the second  field ${\psi}_{+}(x_2^{-}, x^{+})$.

The generalized (kinetic) momenta ${\pi}_{(i)}^{\pm}$ are given by
\begin{equation}
{\pi}_{(i)}^{\pm} = p_{(i)}^{\pm} + e A_{(i),self}^{\pm}
\end{equation}
with
\[
p_{(i)}^{\pm} \equiv i {\partial}_{(i)}^{\pm}
\]
and
\[
A_{(1),self}^{-}(x^{-},x^{+}) \equiv {\phi}_{(1)}^{self}(x^{-},x^{+})=
\]
\[
=\frac{e}{2} \int_0^L dy^{-} \int_0^L dz^{-} {\Phi}^{\star,++}
(y^{-},x^{+}|z^{-},x^{+}) D^{sym}(x^{-},y^{-}|L)
{\Phi}^{++}(y^{-},x^{+}|z^{-},x^{+}),
\]
\[
A_{(2),self}^{-}(x^{-},x^{+}) \equiv {\phi}_{(2)}^{self}(x^{-},x^{+})=
\]
\[
=\frac{e}{2} \int_0^L dy^{-} \int_0^L dz^{-} {\Phi}^{\star,++}
(y^{-},x^{+}|z^{-},x^{+}) D^{sym}(x^{-},z^{-}|L)
 {\Phi}^{++}(y^{-},x^{+}|z^{-},x^{+}),
\]
\[
A_{(i),self}^{+}(x^{-},x^{+}) =0,
\]
the self-potentials ${\phi}_{(i)}^{self}$ being nonlinear integral
expressions.

The self-potential in the mass term
\begin{equation}
{\phi}^{self}(x^{+}) =
\int_0^L dy^{-} \int_0^L dz^{-} \int_0^L d{\eta}^{-}
{\Phi}^{\star,++}(z^{-},x^{+}|y^{-},x^{+})
i{\epsilon}(y^{-} - {\eta}^{-}) {\Phi}^{++}({\eta}^{-},x^{+}|
z^{-},x^{+})
\end{equation}
does not depend on the light-front space coordinate and
does not contribute to the action in the massless case.

We can write the action (3.3) in another, equivalent form.
If ${\phi}_{(i)}^{self}$ are not included into the
generalized momenta and are considered separately, then these
self-potentials can be shown to reduce to the potential
$e^2D^{sym}$. The action becomes
\[
W[{\Phi}]=\frac{1}{2} \int_{- \infty}^{\infty} dx^{+} \int_0^L dx_1^{-}
\int_0^L dx_2^{-} {\Phi}^{\star,++}(x_1^{-},x^{+}|x_2^{-},x^{+})
\cdot (p_{(1)}^{-} + p_{(2)}^{-} +
\]
\begin{equation}
+ e^2D^{sym} - m {\phi}^{self} )
{\Phi}^{++}(x_1^{-},x^{+}|x_2^{-},x^{+}).
\end{equation}
Since there is only one self-potential in the
action (3.6) , it is much simpler to work with this action
rather than with the action (3.3).

In the self-field approach the ${\psi}$-currents are actual
material charge currents, and not just probability currents.
The corresponding charges are actual matter charges. With the
normalization factor (3.2), we get
\[
Q^{+}=e,
\]
i.e. $Q^{+}$ is actual charge of the positive chirality matter.
For our model with a single charge on the circle, the electric
field is not therefore periodic. We must consider at least two
matter fields with charges equal in magnitude and opposite in
sign in order to get vanishing total charge and single-valued
total electric field.

Now we require the action (3.6) to be stationary not
with respect to the variation of the original matter field
${\psi}_{+}(x^{-},x^{+})$ but with respect to the total composite
field only. This is a weaker condition and
leads to the following equation for
${\Phi}^{++}(x_1^{-},x^{+}|x_2^{-},x^{+})$ in the
configuration space :
\begin{equation}
(p_{(1)}^{-} +
p_{(2)}^{-} + e^2D^{sym} - m {\phi}^{self})
{\Phi}^{++}(x_1^{-} , x^{+}|x_2^{-}, x^{+}) =0.
\end{equation}

We next introduce center of mass and relative coordinates according
to
\begin{eqnarray*}
P^{\pm}=p_{(1)}^{\pm} + p_{(2)}^{\pm}  & , & p^{\pm} = p_{(1)}^{\pm}
- p_{(2)}^{\pm} , \\
R= x_{(1)}^{-} + x_{(2)}^{-}  & , & r=x_{(1)}^{-} - x_{(2)}^{-}. \\
\end{eqnarray*}
The configuration space $(r,R)$ is again a torus, but with the circle
length $2L$ $(-L \leq r <L $ , $0 \leq R < 2L)$.
The function $D^{sym}(x_1^{-},x_2^{-}|L)$ can be rewritten as a sum of
center of mass and relative parts :
\[
D^{sym}(x_1^{-},x_2^{-}|L) = D_{-}(r|L) + D_{+}(R|L) ,
\]
\[
D_{-}(r|L) \equiv \frac{1}{2} |r| - \frac{1}{4L} r^2 ,
\]
\[
D_{+}(R|L) \equiv \frac{1}{4L} R^2 .
\]

Eq.(3.7) becomes
\begin{equation}
P^{-}{\Phi}^{++}(r,x^{+}|R,x^{+})=
\{ -e^2(D_{-}(r) + D_{+}(R)) - m {\phi}^{self} \}
{\Phi}^{++}(r,x^{+}|R,x^{+}) .
\end{equation}
Eq.(3.8) is a Hamiltonian form of the bound
state equation. We have in this equation only one time variable conjugate
to the center of mass energy $P^{-}$ ; the relative energy
$p^{-}$ does not enter.

\renewcommand{\ren}{\renewcommand{\theequation}{4.\arabic{equation}}}
\renewcommand{\add}{\addtocounter{equation}{-1}}
\renewcommand{\new}[1]{\renewcommand{\theequation}{4.\arabic{equation}#1}}
\renewcommand{\set}{\setcounter{equation}{0}}

\set

\begin{center}

\section{ANALYSIS OF BOUND STATE EQUATION}

\end{center}

Let us define  the operator
$M^2=P^{+}P^{-}$, so that its eigenvalues correspond to
the invariant mass spectrum of the theory.
By acting on (3.8) by $P^{+}$ we get the Hamiltonian form of
the bound state equation in terms of $M^2$:
\[
M^2 {\Phi}^{++} = \{ -4ie^2 \frac{{\partial}D_{+}(R)}{{\partial}R}
+ 4i (e^2 D_{-}(r) + e^2 D_{+}(R) 
\]
\ren
\begin{equation}
- m{\phi}^{self}) \frac{\partial}
{\partial R} \} {\Phi}^{++}.
\end{equation}

To find eigenfunctions and eigenvalues of $M^2$ we must solve
the equation
\ren
\begin{equation}
M^2 {\Phi}^{++} = {\sigma} {\Phi}^{++},
\end{equation}
where $\sigma$ has the dimension ${< energy >}^2$.

With the normalization condition
\[
\int_{0}^{2L} dR \int_{-L}^{L} dr {\Phi}^{\star,++}(r,R)
{\Phi}^{++}(r,R) =1 ,
\]
Eq.(4.2) is solved by the eigenfunctions
\ren
\begin{equation}
{\Phi}_n^{++}(r,R) = C(m,{\phi}) \frac{f_m(r)}{R^2
+ d_m(r)} \exp\{i \frac{L}{e^2} {\sigma}_n {\cal F}
(d_m(r),R) \},
\end{equation}
where
\[
d_m(r) \equiv - \frac{4L}{e^2} m{\phi}
+ 4L D_{-}(r),
\]
${\phi}$ is a fixed value of the self-potential
${\phi}^{self}$, $f_m(r)$ is an arbitrary function,
$$
{\cal F}(d_m(r),R) = \left \{
\begin{array}{cc}
\frac{1}{\sqrt{d_m(r)}}{\rm arctan}(\frac{R}
{\sqrt{d_m(r)}}) & {\rm for}
\hspace{5 mm} d_m(r)>0, \\
\frac{1}{2\sqrt{|d_m(r)|}} \ln\left|\frac{\sqrt{|d_m(r)|}
- R}{\sqrt{|d_m(r)|} +
 R}\right| & {\rm for}
\hspace{5 mm} d_m(r)<0,
\end{array}
\right.
$$
and
\[
{\cal F}(d_m(r),R) = - \frac{1}{R}
\hspace{5 mm} {\rm for} \hspace{5 mm}
d_m(r)=0.
\]
The normalization constant $C(m,{\phi})$ is
\[
C(m,{\phi}) =\left(\int_0^{2L} dR \int_{-L}^{L} dr
\frac{f_m^2(r)}{(R^2 + d_m(r))^2} \right)^{-1/2}.
\]

The eigenfunctions ${\Phi}_n^{++}$ become singular at those
points of the configuration space $(r,R)$ where
\[
R^2 + d_m(r)=0.
\]
However, if
\begin{equation}
\phi > \frac{5}{4} \frac{e^2L}{m},
\end{equation}
then all singularity points are outside of the torus
$(- L \leq r < L$, $0 \leq R < 2L)$.
There are no singularity points also for negative
values of ${\phi}^{self}$. For any nonzero $m$
we can therefore choose ${\phi}$ in such way that
the condition (4.4) is fulfilled and
the eigenfunctions ${\Phi}_n^{++}$ are not singular.

For $m=0$, the condition (4.4) is not already valid, and
the eigenfunctions are singular at one point
of the configuration space , namely $(r=0$ , $R=0).$

\begin{center}

\subsection{Spectrum}

\end{center}

The spectrum of eigenvalues ${\sigma}_n$ is fixed
by the boundary conditions. From the boundary conditions
for the individual fields ${\psi}_{+}$ we can deduce
in general three boundary conditions for the composite
field ${\Phi}^{++}$ ( see Fig.1) :
\begin{eqnarray*}
{\Phi}^{++}(L|L) & = & \exp\{i2{\pi}{\kappa}_{+}\}{\Phi}^{++}(0|0),\\
{\Phi}^{++}(-L|L) & = & \exp\{i2{\pi}{\kappa}_{+}\}{\Phi}^{++}(0|0),\\
{\Phi}^{++}(0|2L) & = & \exp\{i4{\pi}{\kappa}_{+}\}{\Phi}^{++}(0|0).\\
\end{eqnarray*}
Nevertheless, only part of them are really valid.
Indeed, the preexponential in the solution (4.3) is not a
constant and depends on the relative coordinate $r$. Only
those boundary points at which the values of the
preexponential coincide can be used in the boundary
conditions.
Let us first put $f_m(r)=1$. Then the preexponential
$\frac{1}{R^2+d_m(r)}$ takes the same value at two
boundary points $(-L,L)$ and $(L,L)$. So we have one
boundary condition (see Fig.2a)
\ren
\begin{equation}
{\Phi}^{++}(-L,L) = {\Phi}^{++}(L,L)
\end{equation}
which can be considered as an equation for $\sigma$, while
${\kappa}_{+}$ remains arbitrary.

For the solution (4.3) with $f_m(r)=1$, from Eq.(4.5) we get
\[
\frac{L}{e^2} {\sigma}_n [ {\cal F}(d_m(-L),L) -
{\cal F}(d_m(L),L) ] =2{\pi}n, \hspace{5 mm} n \in \cal Z.
\]
Since $d_m(-r)=d_m(r)$, the expression in the squared
brackets vanishes and ${\sigma}_n$ drops out of the
boundary conditions. In the case $f_m(r)=1$ we therefore
can not derive any information concerning the spectrum.

For the general case $f_m(r) \neq 1$, we can use one boundary
point more in the boundary conditions. If we take $f_m(r)$
with the boundary values connected as
\[
f_m(-L)=f_m(L),
\]
\[
\frac{f_m(L)}{f_m(0)} = \frac{L^2 + d_m(L)}{4L^2 + d_m(0)},
\]
then the preexponential $\frac{f_m(r)}{R^2 + d_m(r)}$ takes
the same value at three boundary points $(-L,L)$, $(L,L)$, and
$(0,2L)$. The corresponding boundary conditions are
(see Fig. 2b)
\begin{eqnarray*}
{\Phi}^{++}(0|2L) & = & \exp\{i2{\pi}{\kappa}_{+}\}
{\Phi}^{++}(L|L),\\
{\Phi}^{++}(0|2L) & = & \exp\{i2{\pi}{\kappa}_{+}\}
{\Phi}^{++}(-L|L),\\
{\Phi}^{++}(L|L) & = & {\Phi}^{++}(-L|L).\\
\end{eqnarray*}
Again ${\sigma}_n$ drops out of the third boundary condition,
while the first two ones coincide and give us for
${\kappa}_{+}=0$ the spectrum
\begin{equation}
{\sigma}_n = \frac{2{\pi}e^2}{{\alpha}_m} n, \hspace{5 mm}
n \in \cal Z,
\end{equation}
which is linear in $n$ , where
\[
{\alpha}_m \equiv L [ {\cal F}(d_m(0),2L) -
{\cal F}(d_m(L),L) ],
\]
and $n$ must be taken positive for ${\alpha}_m>0$ and
negative for ${\alpha}_m<0$.

Eqs.(4.3) and (4.6) represent a solution of the invariant
mass squared operator eigenvalue problem for a fixed value
of ${\phi}^{self}$.

\begin{center}

\subsection{Massless case}

\end{center}

For the massless case, there are no self-energy terms in
the bound state equation. The eigenfunctions of $M^2$
become
\begin{equation}
{\Phi}^{(0),++}_n = C(0,0) \frac{f_0(r)}{R^2 + d_0^{(0)}(r)}
\exp\{i \frac{L}{e^2} {\sigma}_n^{(0)} {\cal F}^{(0)}
(d_0^{(0)}(r),R) \},
\end{equation}
where $d_0^{(0)}(r)=4LD_{-}(r)$ is positive for all nonzero
$r$ and vanishes at $r=0$. The superscript $(0)$ means that
(4.7) is the solution of the invariant mass squared operator
eigenvalue problem without the self-energy terms.

We easily evaluate ${\alpha}_0^{(0)}$ as
\[
{\alpha}_0^{(0)} = - \frac{1}{2} (1 + \frac{\pi}{2})<0,
\]
so that the spectrum takes the form
\begin{equation}
{\sigma}_n^{(0)} = \frac{8{\pi}e^2}{{\pi} +2} n,
\hspace{5 mm} n=0,1,2,...   .
\end{equation}

For the boundary values of $f_0(r)$ we have
\[
f_0(-L) = f_0(L),
\]
\[
f_0(L) = \frac{1}{2} f_0(0).
\]
One possible choice of $f_0(r)$ is
\[
f_0(r) = {\theta}(\frac{L^2}{4} - r^2) + \frac{1}{2}
{\theta}(r^2 - \frac{L^2}{4})
\]
(see Fig. 3). With this choice the normalization constant
takes the value
\[
C(0,0) = L (\frac{1}{4} \ln{\frac{3}{2}} +
\ln{\frac{11}{7}})^{-1/2} \approx 1,35 \cdot L .
\]

\begin{center}

\section{DISCUSSION}

\end{center}

1. In our study of the bound state problem in two-dimensional
$QED$ we have tried to combine the advantages of both the
self-field approach and the light-front formulation. The
self-field approach allows us to construct a relativistic
bound state equation. Bound states are described by a composite
matter field which is a bilinear combination of the original
matter field and therefore $4$-component. On light front only
one of these components is non-zero, so the bound state
equation is simply one-component.

We have derived the Hamiltonian form of the bound state
equation for the invariant mass squared operator. The
equation includes a self-potential which enters the mass
term. For a fixed value of the self-potential, we have
solved the eigenvalue problem of the invariant mass
squared operator and found its spectrum as well as
the corresponding eigenfunctions. The invariant mass
spectrum turns out to be discrete and linear. For the
massless matter fields, when the self-potential contribution
vanishes , we have evaluated the
invariant mass spectrum explicitly.

2. Our solution of the invariant mass squared operator
eigenvalue problem is not complete. In particular, in
the massless case the eigenfunctions are singular at one
of the boundary points, so only a part of the boundary
conditions can be employed . Since just the boundary
conditions determine the spectrum, some information about
the spectrum is lost. As a result, the invariant mass spectrum
obtained in our work differs from the well-known one
of the second-quantized massless $QED_2$ on light
front ( the light-front Schwinger model) \cite{berg77}.

There are several puzzles or problems regarding the
light-front formulation, e.g. the null plane and missing
degrees of freedom, causality and boundary conditions,
the zero mass limit and so on (~see, for example, \cite{rob96}).
Many aspects of the Schwinger model on light front such
as the anomaly and the $\theta$-vacuum  can not be
understood without worrying about these problems \cite{mc91}.
Ignoring of the problems works only in one case when
we evaluate the spectrum. The spectrum of physical
bosons of the light-front Schwinger model is reproduced
exactly without a complete formulation in which all
these problems are solved.

In the self-field approach we first find the
eigenfunctions of the two-body Hamiltonian or the
invariant mass squared operator, and then impose the
boundary conditions and determine the spectrum. An
exact expression for the spectrum can not be therefore
reproduced unless a complete construction of the
two-body Hamiltonian on light front is given.

In the usual equal-time formulation of the self-field QED
we have only one time variable in the relativistic
bound state equation \cite{barut90,barsa94}.
We could start in principle
with the field
\[
{\Phi}(x_{(1)}^1,x_{(1)}^0|x_{(2)}^1,x_{(2)}^0) =
{\psi}_1(x_{(1)}^1,x_{(1)}^0) \otimes
{\psi}_2(x_{(2)}^1,x_{(2)}^0)
\]
composed of the matter fields ${\psi}_1$ and
${\psi}_2$ taken at different times.
However, only the center of mass energy conjugate
to the time $t \equiv x_{(1)}^0 + x_{(2)}^0$
enters the bound state equation and contributes
to the two-body Hamiltonian. The relative energy
conjugate to the relative time $\tau \equiv
x_{(1)}^0 - x_{(2)}^0$ drops out of this equation
automatically. That is why, without loss of
generality, we put $x_{(1)}^0=x_{(2)}^0$
from the beginning.

In the light-front formulation the situation seems
to be quite different. The time and spatial
variables are mixed, so both the light-front
center of mass and relative energies can
contribute to the bound state equation. In the
present work, following the equal-time formulation
prescription we have taken the matter fields in
the ansatz (3.1) at the same light-front time,
and the relative energy contribution was lost.
A generalization of the ansatz (3.1) is then
obvious: we should take the matter fields at different
light-front times.

As shown in \cite{rob96}, infrared regularization using a finite
volume and a careful treatment of the boundary surfaces
are required to construct a light-front theory that is
equivalent to the equal-time theory. The Hamiltonian
and other conserved charges obtained in this way are
guaranted to be identical to the ones we would construct
in the equal-time formulation. We believe that with the
finite box regularization and the generalized ansatz for
the composite matter field it will be possible to perform
a complete construction of the light-front two-body
Hamiltonian in our model. If so, this will allow us to
derive exact and complete expressions for the invariant
mass spectrum and the eigenfunctions.

3. For non-zero $m$, the solution (4.3) is a formal one,
because it contains an undetermined value of the
self-field potential.

In the self-field quantum electrodynamics
the self-field potentials are calculated by iteration
procedure.
To lowest order of iteration we solve the
Hamiltonian eigenvalue problem  without   the
self-energy   terms $({\phi}_{(0)}^{self} \equiv
{\phi}_{0}=0 )$. Next we substitute the eigenfunctions
obtained into the expressions for the self-field
potentials and calculate these potentials explicitly.
For our model with ${\phi}^{self}$ independent of
spatial coordinate, we would get that ${\phi}_{(1)}^{self}
\equiv {\phi}_1$ is simply a number.

To the next
order of iteration we find the solution of the
eigenvalue problem already with the potential
${\phi}_1$.  Using    the new  eigenfunctions, we
calculate ${\phi}_{(2)}^{self} \equiv {\phi}_2$, then
find the eigenfunctions and eigenvalues corresponding
to ${\phi}_2$ and so on, the potentials ${\phi}_2$ ,
${\phi}_3$ , ... depending on $m$.

If there is a small parameter in the theory, then
we can often stop the iteration procedure already
after the first order. In our model we could take
the mass
$m$   as such a parameter and consider the mass
contribution to the bound state eigenfunctions and
eigenvalues as small corrections to the corresponding
eigenfunctions and eigenvalues for the vanishing
mass. To do actual calculations in the massive case
we therefore need again the complete solution of
the eigenvalue problem for the massless case.
The mass corrections in the second-quantized light-front
$QED_2$ were calculated in \cite{hara97}.

\vspace{2 cm}

\begin{center}

{\bf ACKNOWLEDGMENT}

\end{center}

This research was supported by Deutscher Akademischer
Austauschdienst and done partially during the visit
of the author to the Max-Planck-Institut f\"ur Kernphysik,
Heidelberg.

\newpage

\begin{center}
{\large \bf Figure Captions}
\end{center}

\vspace{1 cm}

{\bf FIG. 1.} A schematic representation of the boundary
conditions for the composite matter field on the configuration
space $(-L \leq r < L$, $0 \leq R < 2L)$. Each two
boundary points connected by a dotted line are related also
by the corresponding boundary condition.

\vspace{2 cm}

{\bf FIG. 2.} A schematic representation of the boundary
conditions for the solution (4.3). \\ (a): $f_m(r)=1$ ,
(b): $f_m(r) \neq 1$.

\vspace{2 cm}

{\bf FIG. 3.} A special choice of the function $f_0(r)$:
$f_0(r) = \theta(\frac{L^2}{4} - r^2) + \frac{1}{2}
\theta(r^2 - \frac{L^2}{4})$.

\newpage

\vspace{3 cm}

\setlength{\unitlength}{1cm}

\begin{center}
\begin{picture}(12,7)(-6,0)
\thinlines
\put (0,-.1){\makebox(0,0)[tl]{(0,0)}}
\put (-3,-.1){\makebox(0,0)[tr]{-L}}
\put (3,-.1){\makebox(0,0)[tl]{L}}
\put (.1,6.1){(0,2L)}
\put (-3.1,3){\makebox(0,0)[tr]{(-L,L)}}
\put (3.1,3){\makebox(0,0)[tl]{(L,L)}}
{\thinlines
\put (-4,0){\line(1,0){4}}
\put (0,0){\line(1,0){3}}
\put (3,0){\vector(1,0){1}}
\put (0,6){\vector(0,1){1}}}
{\thinlines
\put (-3,0){\line(0,1){6}}
\put (3,0){\line(0,1){6}}
\put (-3,6){\line(1,0){6}}}
\put (4,.2){r}
\put (.2,7){R}
\put (0,0){\circle*{.2}}
\put (-3,3){\circle*{.2}}
\put (3,3){\circle*{.2}}
\put (0,6){\circle*{.2}}
{\thicklines
\multiput (0,0)(.3,.3){10}{\circle*{.1}}
\multiput (0,0)(0,.3){20}{\circle*{.1}}
\multiput (0,0)(-.3,.3){10}{\circle*{.1}}}

\end{picture}

\vspace{2 cm}

{\bf FIG. 1.}

\end{center}

\newpage

\vspace{3 cm}

\begin{center}
\begin{picture}(8,7)(-4,0)
\thinlines
\put (0,-.1){\makebox(0,0)[tl]{(0,0)}}
\put (-3,-.1){\makebox(0,0)[tr]{-L}}
\put (3,-.1){\makebox(0,0)[tl]{L}}
\put (.1,6.1){(0,2L)}
\put (-3.1,3){\makebox(0,0)[tr]{(-L,L)}}
\put (3.1,3){\makebox(0,0)[tl]{(L,L)}}
{\thinlines
\put (-4,0){\line(1,0){4}}
\put (0,0){\line(1,0){3}}
\put (3,0){\vector(1,0){1}}
\put (0,0){\line(0,1){6}}
\put (0,6){\vector(0,1){1}}}
{\thinlines
\put (-3,0){\line(0,1){6}}
\put (3,0){\line(0,1){6}}
\put (-3,6){\line(1,0){6}}}
\put (0,0){\circle*{.2}}
\put (-3,3){\circle*{.2}}
\put (3,3){\circle*{.2}}
\put (0,6){\circle*{.2}}
\put (4,.2){r}
\put (.2,7){R}
{\thicklines
\multiput (-3,3)(.3,0){20}{\circle*{.1}}}

\end{picture}

\vspace{1 cm}

(a)

\end{center}

\vspace{1 cm}

\begin{center}
\begin{picture}(8,7)(-4,0)
\thinlines
\put (0,-.1){\makebox(0,0)[tl]{(0,0)}}
\put (-3,-.1){\makebox(0,0)[tr]{-L}}
\put (3,-.1){\makebox(0,0)[tl]{L}}
\put (.1,6.1){(0,2L)}
\put (-3.1,3){\makebox(0,0)[tr]{(-L,L)}}
\put (3.1,3){\makebox(0,0)[tl]{(L,L)}}
{\thinlines
\put (-4,0){\line(1,0){4}}
\put (0,0){\line(1,0){3}}
\put (3,0){\vector(1,0){1}}
\put (0,0){\line(0,1){6}}
\put (0,6){\vector(0,1){1}}}
{\thinlines
\put (-3,0){\line(0,1){6}}
\put (3,0){\line(0,1){6}}
\put (-3,6){\line(1,0){6}}}
\put (4,.2){r}
\put (.2,7){R}
\put (0,0){\circle*{.2}}
\put (-3,3){\circle*{.2}}
\put (0,6){\circle*{.2}}
\put (3,3){\circle*{.2}}
{\thicklines
\multiput (-3,3)(.3,.3){10}{\circle*{.1}}
\multiput (-3,3)(.3,0){20}{\circle*{.1}}
\multiput (0,6)(.3,-.3){10}{\circle*{.1}}}
\end{picture}

\vspace{1 cm}

(b)

\vspace{1 cm}

{\bf FIG. 2.}

\end{center}

\newpage

\vspace{3 cm}

\setlength{\unitlength}{1cm}
\begin{center}
\begin{picture}(8,5)(-4,0)
\thinlines
\put (-1.5,-.1){\makebox(0,0)[tr]{-$\frac{1}{2}$L}}
\put (-3,-.1){\makebox(0,0)[tr]{-L}}
\put (1.5,-.1){\makebox(0,0)[tl]{$\frac{1}{2}$L}}
\put (3,-.1){\makebox(0,0)[tl]{L}}
\put (.1,2){$\frac{1}{2}$}
\put (.1,4.1){1}
{\thinlines
\put (-4,0){\line(1,0){4}}
\put (0,0){\line(1,0){3}}
\put (3,0){\vector(1,0){1}}
\put (0,0){\line(0,1){4}}
\put (0,4){\vector(0,1){1}}}
\put (0,2){\circle*{.1}}
\put (0,4){\circle*{.1}}
\put (-1.5,0){\circle*{.1}}
\put (-3,0){\circle*{.1}}
\put (0,0){\circle*{.1}}
\put (1.5,0){\circle*{.1}}
\put (3,0){\circle*{.1}}
\put (0,-.1){\makebox(0,0)[tl]{0}}
{\thinlines
\multiput (-1.5,0)(0,.4){10}{\line(0,1){.2}}
\multiput (3,0)(0,.4){5}{\line(0,1){.2}}
\multiput (-3,0)(0,.4){5}{\line(0,1){.2}}
\multiput (1.5,0)(0,.4){10}{\line(0,1){.2}}}
{\thicklines
\put (-1.5,4){\line(1,0){3}}
\put (-3,2){\line(1,0){1.5}}
\put (1.5,2){\line(1,0){1.5}}}
\thinlines
\put (4,.2){r}
\put (.2,5){$f_0$}
\end{picture}

\vspace{2 cm}

{\bf FIG. 3.}

\end{center}

\end{document}